# Properties and Collapse of the Ferromagnetism in UCo$_{1-x}$Ru$_x$Al Studied in Single Crystals

Jiří Pospíšil[1,2*], Petr Opletal[1], Michal Vališka[1], Yo Tokunaga[2], Anne Stunault[3], Yoshinori Haga[2], Naoyuki Tateiwa[2], Béatrice Gillon[4], Fuminori Honda[5], Tomoo Yamamura[6], Vojtěch Nižňanský[1], Etsuji Yamamoto[2], and Dai Aoki[5]

[1] Charles University in Prague, Faculty of Mathematics and Physics, Department of Condensed Matter Physics, Ke Karlovu 5, Prague 2, Czech Republic
[2] Advanced Science Research Center, Japan Atomic Energy Agency, Tokai, Ibaraki, 319-1195, Japan
[3] Institut Laue Langevin, 71 Avenue des Martyrs, CS 20156, F-38042 Grenoble Cedex 9, France
[4] Laboratoire Léon Brillouin, UMR12 CEA-CNRS, Bât 563, CEA Saclay, 91191 Gif sur Yvette Cedex, France
[5] Institute for Materials Research, Tohoku University, Oarai, Ibaraki 311-1313, Japan
[6] Institute for Materials Research, Tohoku University, 2-1-1, Katahira, Aoba, Sendai, Miyagi 980-8577, Japan

We have investigated the ferromagnetic (FM) phase which suddenly develops in UCo$_{1-x}$Ru$_x$Al and is isolated by paramagnetic regions on both sides from the parent UCoAl and URuAl. For that purpose we have grown high quality single crystals with $x$ = 0.62, 0.70, 0.74, 0.75 and 0.78. The properties of the FM phase have been investigated by microscopic and macroscopic methods. Polarized neutron diffraction on a single crystal with $x$ = 0.62 revealed the gradual growth of the hybridization between U and $T$-site in the U – $T$ plane with increasing $x$. Hybridization works here as a mediator of the strong indirect interaction, while the delocalized character of the 5$f$ states is still conserved. As a result very weak spontaneous magnetic moments are observed for all alloys with magnitude nearly in proportion to the $T_\mathrm{C}$ for $x$ < 0.62, while an enormous disproportion exists between them near $x_\mathrm{crit.}$. The magnetization, specific heat, electrical resistivity, and Hall effect measurements confirmed that the FM transition is suppressed continuously at the critical concentration $x_\mathrm{crit.} \approx 0.77$. Two quantum critical points are then expected on both sides of the FM dome. We propose a scenario that the order of the FM/PM transition differs at opposite boundaries of the FM dome. We conclude that both criticalities are influenced by disorder. Criticality on the UCoAl side has the character of a clean FM metal, while on the Ru rich side it has the character of a magnetically inhomogeneous system involving a Griffiths phase.

E-mail: jiri.pospisil@centrum.cz
       pospisil.jiri@jaea.go.jp

# 1 Introduction

The physics of uranium 5$f$ electron systems is a subject of continuing interest because many uranium compounds reveal strong electron correlations leading to phenomena like non-Fermi Liquid (NFL) behavior or unconventional superconductivity.

In the U$TX$ series recent interest focuses on two ground state paramagnets (PM) UCoAl[1-6] and URuAl[7] with features differing significantly from ordinary PMs. UCoAl has attracted because of the very low critical field of the metamagnetic transition ($B_c \approx 0.6$ T), which occurs only in fields parallel to the $c$ axis[8]. This Ising-type anisotropy arises from ferromagnetic fluctuations which lack any transversal component[9, 10]. The paramagnetic ground state of URuAl is also unusual. Magnetic susceptibility is characterized by a broad maximum around 50 K and band structure calculations suggest an anomalous cancelation of the U orbital $\mu_L^U$- and spin $\mu_S^U$- momentum finally responsible for the lack of magnetic ordering[7]. Despite PM ground state, magnetocrystalline anisotropy is conserved in URuAl in contrast to the true Pauli PM UFeAl, which is magnetically isotropic[11].

A spectacular feature of the UCo$_{1-x}$Ru$_x$Al[12] solid solutions, the subject of our research, is that they do not reflect the magnetic ground states of the parent compounds. Instead, a robust FM phase emerges isolated on both sides by a PM phase. Already a very small concentration of Ru ($x \approx 0.01$) in UCoAl suddenly gives rise to stable FM with an enormous $T_C$ jump from zero up to almost 20 K[13, 14]. Further substitution by Ru increases $T_C$ to 60 K for $x \approx 0.3$. Beyond this optimum amount, the $T_C$ is suppressed and survives up to the critical Ru concentration $x_{crit.} \approx 0.8$. The observed sudden FM in UCoAl is a surprisingly general phenomenon for many substituting $T$-metals. On the other hand, the later FM phase behavior significantly changes at higher concentrations[13, 15-18]. The differing features of the FM phase at opposite boundaries indicate evolution of the magnetic interactions through the system. So far, however, the microscopic relation of the strong FM in UCo$_{1-x}$Ru$_x$Al to particular properties of the parent PM compounds is unclear.

To explore the above issue, we have investigated a series of UCo$_{1-x}$Ru$_x$Al single crystals with compositions $0.62 \leq x \leq 0.78$ by magnetization, heat capacity, electrical resistivity, Hall effect and especially by polarized neutron diffraction (PND). This is the first PND study in alloyed ZrNiAl-type U$TX$ system.

Our research opens the question of quantum criticality. Since the FM/PM boundaries are expected to be different, the criticality should also be different. Much scientific effort has already been spent on resolving the features of the metamagnetic-ferromagnetic transition on the UCoAl side[14, 19-22]. NMR studies show that the quantum critical point (QCP) is not reached because the transition becomes first order. Nevertheless, the existence of a tri-critical point (TCP) connected with the development of the strong spin fluctuations was confirmed[22]. On the other hand, criticality on the Ru-rich side is an entirely untouched subject. Other U$TX$ alloying systems with TiNiSi-type structure are currently intensively studied and a non-Fermi liquid state appears at the substituent-rich boundaries of the similarly emergent magnetic phases[23-26]. In contrast, the FM phase in UCo$_{1-x}$Ru$_x$Al reaches much higher $T_C$, survives to significantly higher concentrations of Ru and, in particular, the FM/PM transition is predicted not to be a simple one[12].

While first order transitions are predicted for the clean FM metals, second order FM/PM transitions can extend down to 0 K in disordered metals[27]. Thus, UCo$_{1-x}$Ru$_x$Al represents a unique system for research where a FM QCP can appear in both limits of substitutional disorder – in the limit of a clean FM metal (UCoAl side of the FM dome) and disordered FM metal (URuAl side of the FM dome).

## 2 Experimental details

UCo$_{1-x}$Ru$_x$Al single crystals with nominal compositions $x$ = 0.60, 0.73 and 0.81 were grown by Czochralski method in tri-arc and tetra-arc furnaces from polycrystalline precursors. The obtained single crystals were cylindrical with a typical length of 30-50 mm and a diameter of 2 – 4 mm. The quality was checked by X-ray Laue method with a photosensitive image plate detector. All the single crystals were wrapped in a tantalum foil (3N), sealed in quartz tubes in 10$^{-6}$ mbar vacuum, annealed for 7 days at 900°C and then cooled down slowly to prevent internal stress. A precise spark erosion saw was used to cut appropriately shaped samples for all the planned experiments. Structural characterization was performed by X-ray powder diffraction using a BRUKER D8 Advance diffractometer. The recorded X-ray powder patterns were evaluated using Rietveld analysis implemented in the FullProf software[28-30]. Phase analysis and test of the stoichiometry of the single crystals from the tri-arc furnace were performed using Tescan Mira I LMH equipped with a Bruker AXS energy dispersive X-ray detector (EDX). Single crystals prepared in the tetra-arc furnace were analyzed using an electron probe microanalyzer EPMA JXA-8230 (JEOL).

The temperature dependent magnetization, AC susceptibility and magnetization curves were measured along the principal crystallographic directions down to $T$ = 1.8 K in applied magnetic fields up to 7 T using a commercial magnetometer MPMS 7T (Quantum Design). The electrical resistivity and heat capacity measurements were carried out down to 1.8 K (0.4 K using $^3$He insert) with applied magnetic fields up to 14 T using a commercial physical property measurement system PPMS 14 T (9 T) (Quantum Design). The Hall Effect was measured down to 20 mK in a 14 T magnetic field in an Oxford Instruments dilution refrigerator with the sample stage mounted directly inside the mixing chamber.

Crystal structure of the $x$ = 0.62 single crystal was determined by neutron diffraction on the 5C2 4-circle diffractometer at 50 K in LLB, Saclay. The monochromatic neutron beam of wavelength 0.83Å was obtained from a Cu monochromator (220 plane). The polarized neutron diffraction experiments were performed on the same sample on the 5C1 diffractometer at the same institution. On 5C1, the polarized 0.84 Å monochromatic neutron beam was obtained using a Cu$_2$MnAl Heusler monochromator (111 plane). A Position Sensitive Detector (64 vertical $^3$He tubes) was used for neutron detection.

## 3 Results

### 3.1 Structure and composition analysis

High quality of the each single crystal was verified by the Laue method, as evidenced by well separated sharp reflections. A small piece of each single crystal was pulverized to a fine powder for X-ray powder diffraction (XRPD) at room temperature. The Rietveld analysis of the XRPD patterns confirmed the hexagonal ZrNiAl-type structure and the non-linear evolution of the lattice parameters as a function of $x$, in agreement with the original work[12].

The concentrations of all single crystals were determined by either EPMA chemical analysis or EDX. The results are summarized in Table I. In the following, we will refer to the samples by their real concentrations. It is important to mention that no sign of any concentration gradient was found in the slices cut perpendicularly to the growth axis. All subsequent measurements were done using samples from zones as narrow as possible to eliminate the effect of the weak concentration gradients along the single crystals ingots. No spurious phases were detected in any of the single crystals.

TABLE I. Comparison of the nominal and real concentrations for all the single crystals.

| Nominal concentration | Real concentration | Furnace type | Method |
|---|---|---|---|
| UCo$_{0.19}$Ru$_{0.81}$Al | UCo$_{0.22}$Ru$_{0.78}$Al | tri-arc | EDX |
| UCo$_{0.27}$Ru$_{0.73}$Al | UCo$_{0.25}$Ru$_{0.75}$Al (bottom part) | tetra-arc | EPMA |
|  | UCo$_{0.26}$Ru$_{0.74}$Al (top part) | tetra-arc | EPMA |
| UCo$_{0.27}$Ru$_{0.73}$Al | UCo$_{0.30}$Ru$_{0.70}$Al | tri-arc | EDX |
| UCo$_{0.40}$Ru$_{0.60}$Al | UCo$_{0.38}$Ru$_{0.62}$Al | tetra-arc | EPMA |

In the neutron diffraction experiment we have accurately determined the crystal structure of UCo$_{0.38}$Ru$_{0.62}$Al single crystal from the measurement of 177 inequivalent reflections using the 5C2 4-circle diffractometer. The data were analyzed using Fullprof software package and the resulting structure is given in Table II.

TABLE II. Evaluated crystal structure of the UCo$_{0.38}$Ru$_{0.62}$Al. Coefficients of the extinction parameter q(hkl) = $q_1 h^2 + q_2 k^2 + q_3 l^3 + q_4 hk + q_5 hl + q_6 kl$ are listed as well. Lattice parameters are in agreement with PXRD listed later in conclusions.

| Site | X | Y | Z | Occ. |
|---|---|---|---|---|
| U 3g | 0.5834(1) | 0 | 1/2 | 1 |
| Co$_1$ 1b | 0 | 0 | 1/2 | 0.725(16) |
| Ru$_1$ 1b | 0 | 0 | 1/2 | 0.274(16) |
| Co$_2$ 2c | 1/3 | 2/3 | 0 | 0.226(13) |
| Ru$_2$ 2c | 1/3 | 2/3 | 0 | 0.774(13) |
| Al 3f | 0.2338(3) | 0 | 0 | 1 |
| Displacement temp. factors | $\beta_{11}$ | $\beta_{22}$ | $\beta_{33}$ | $\beta_{12}$ |
| U 3g | 0.17(2) | 0.22(2) | 0.16(2) | 0.05(1) |
| Co$_1$ 1b | 0.32(7) | 0.32(7) | 0.39(9) | -0.04(4) |
| Ru$_1$ 1b | 0.32(7) | 0.32(7) | 0.39(9) | -0.04(4) |
| Co$_2$ 2c | 0.16(3) | 0.16(3) | 0.29(4) | -0.12(2) |
| Ru$_2$ 2c | 0.16(3) | 0.16(3) | 0.29(4) | -0.12(2) |
| Al 3f | 0.19(4) | 0.28(5) | 0.43(5) | -0.01(3) |
| Extinction correction | $q_1$=0.45(4) | $q_2$=0.54(1) | $q_3$=0.450(2) | $q_{4,5,6}$=0 |

### 3.2 Magnetic properties

All the magnetization data were collected with the magnetic field applied along the c-axis. The temperature dependences of the magnetizations are summarized in Fig. 1. The single crystal with $x = 0.78$ does not reveal any sign of magnetic order down to 1.8 K. A clear rise of the magnetization at low temperature was detected for all the other compositions. Field cooled (FC) and zero field cooled (ZFC) magnetizations reveal clearly different behavior when a rapid drop is observed in ZFC branches. The $T_C$ values were estimated as the temperatures of the first derivative $\partial M/\partial T$ maxima of the FC curves (Fig. 1) and by Arrott plot analysis[31]. All the results, listed in Table III, confirm the gradual drop of $T_C$ with increasing Ru content $x$. Extrapolating $T_C$ to higher $x$ shows that $T_C$ should vanish for $x \approx 0.77$, which is consistent with our observation in UCo$_{0.22}$Ru$_{0.78}$Al.

Representative examples of the Arrott plots are displayed in Fig. 2. UCo$_{0.38}$Ru$_{0.62}$Al magnetization isotherms of $M^2$ vs. $\mu_0 H/M$ are linear showing mean-field behavior[32]. Other compounds with higher Ru content gradually deviate from this linear trend. Magnetizations can be described by parallel isotherms in a modified Arrott plot [$M^{(1/\beta)} = (\mu_0 H/M)^{(1/\gamma)}$] where $\beta$ and $\gamma$ are universality class critical exponents[33]. $\beta$ does not change with $x$. On the other hand, $\gamma$ develops rapidly; 0.8, 0.6, 0.55 for $x$ = 0.70, 0.74 and 0.75, respectively.

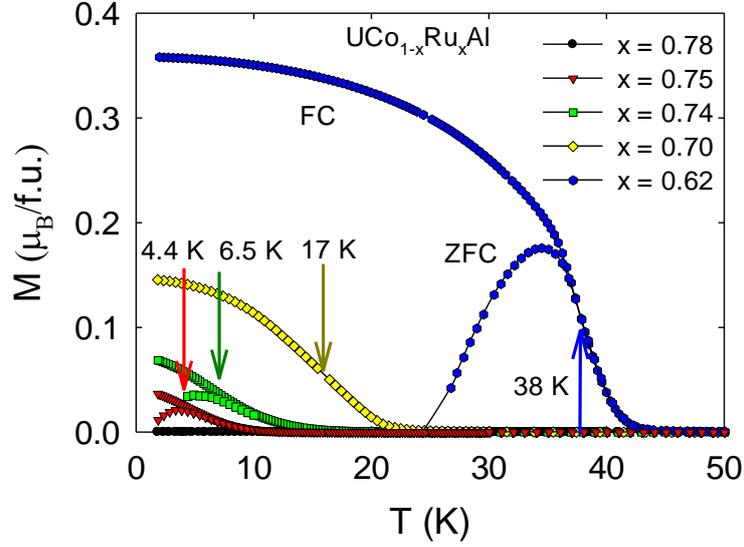

FIG. 1. (Color online) Temperature dependence of the FC and ZFC magnetization in magnetic field of 10 mT applied along the c-axis. The arrows mark the estimated $T_C$ obtained using the first derivatives ($\partial M/\partial T$) of the magnetization curves.

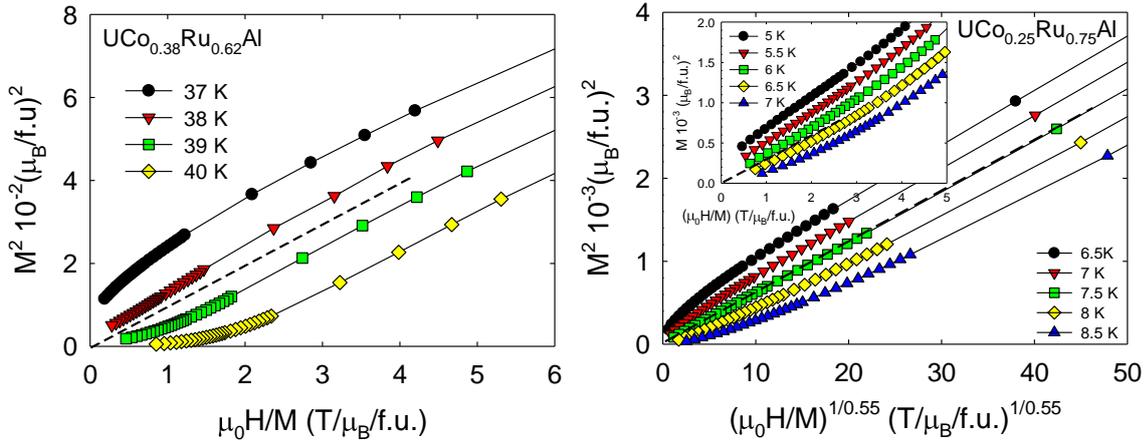

FIG. 2. (Color online) Arrott plots constructed from the magnetization isotherms of $UCo_{0.38}Ru_{0.62}Al$ and $UCo_{0.25}Ru_{0.75}Al$. Estimated $T_C$ are shown by dashed lines. The inset shows $UCo_{0.25}Ru_{0.75}$ isotherms in mean field representation showing deviation from the linear trend.

TABLE III. The $T_C$ of all compounds was estimated using Arrott plot (*modified Arrott plot) analysis, maximum of the first derivative $\partial M/\partial T$, position of the cusp anomaly in electrical resistivity data ($\rho$) and inflection points of the anomaly in heat capacity data (HC), respectively. Temperatures where $\mu_{spont.}$ and hysteresis appear are also tabulated. The saturation moments are taken as the magnetization at 1.8 K in a 7 T magnetic field. The spontaneous magnetizations were obtained from the hysteresis loops by extrapolating to zero field at 1.8 K. The effective magnetic moments ($\mu_{eff.}$) were evaluated from the paramagnetic part of the reciprocal susceptibility together with $\chi_0$ using a modified Curie-Weiss law; the Sommerfeld $\gamma$ coefficients were extrapolated from the $C_p/T$ vs. $T^2$ data. The critical exponents of the electrical resistivity $\rho$ were evaluated using the exponential function $\rho = \rho_0 + AT^x$.

| UCo$_{1-x}$Ru$_x$Al | $x = 0.62$ | $x = 0.70$ | $x = 0.74$ | $x = 0.75$ | $x = 0.78$ |
|---|---|---|---|---|---|
| $T_C$ [K] Arrott | 38.5 | 18* | 10.5* | 7.5* | - |
| $T_C$ [K] M ($\partial M/\partial T$) | 38 | 17 | 6.5 | 4.5 | - |
| $T$ [K] $\mu_{spont.}$ | 39 | | 13 | 9 | - |
| $T$ [K] hysteresis | 35 | | 6.5 | 4.5 | - |
| $T_C$ [K] $\rho$ | 39 | - | - | - | - |
| $T_C$ [K] HC | 35 | - | - | - | - |
| $\mu_{spont.}$ [$\mu_B$/f.u.] | 0.44 | 0.15 | 0.07 | 0.04 | 0 |
| $\mu_{sat.}$ [$\mu_B$/f.u.] (7 T) | 0.48 | 0.27 | 0.23 | 0.18 | 0.10 |
| $\mu_{eff.}$ [$\mu_B$/f.u.] | 1.9 | 2.2 | 2.0 | 1.9 | 2.0 |
| $\chi_0$ [$10^{-9}$ m$^3$/mol] | 10.4 | 18.9 | 9.0 | 8.0 | 10.3 |
| $\theta_p$ [K] | 39.5 | 19.1 | 9.1 | 7.3 | 1.2 |
| $\gamma$ [J/molK$^2$] | 55 | 70 | 65 | 65 | 70 |
| $\rho$ | $\sim T^{2.0}$ | $\sim T^{1.7}$ | $\sim T^{1.5}$ | $\sim T^{1.4}$ | $\sim T^{1.6}$ |

FM transitions were confirmed by AC susceptibility (Fig. 3). As expected, all compounds reveal peaks except UCo$_{0.22}$Ru$_{0.78}$Al. The peaks are very weak in the case of UCo$_{0.25}$Ru$_{0.75}$Al and UCo$_{0.26}$Ru$_{0.74}$Al, almost 10-times lower than UCo$_{0.38}$Ru$_{0.62}$Al and rather broad. Detail analysis showed that the wide peaks consist of two peaks with maxima separated by ~ 3 K and gap which does not change significantly with increasing $x$. This is quite different behavior than that of the polycrystalline samples with two very well-separated maxima, where separation grows as the function of $x$ and reaches almost 20 K near $x_{crit.}$[12]. Nevertheless the relative size of the peaks changes with increasing $x$. The higher temperature one starts to dominate approaching $x_{crit.}$. Another specific feature is their strong sensitivity to the modulation field frequency which causes the peaks to be suppressed at higher frequency (Fig. 3). The frequency dependence appears dominantly around the high temperature peaks for $x > 0.7$. The response of the polycrystalline sample AC susceptibility to the modulation field frequency was not tested[12].

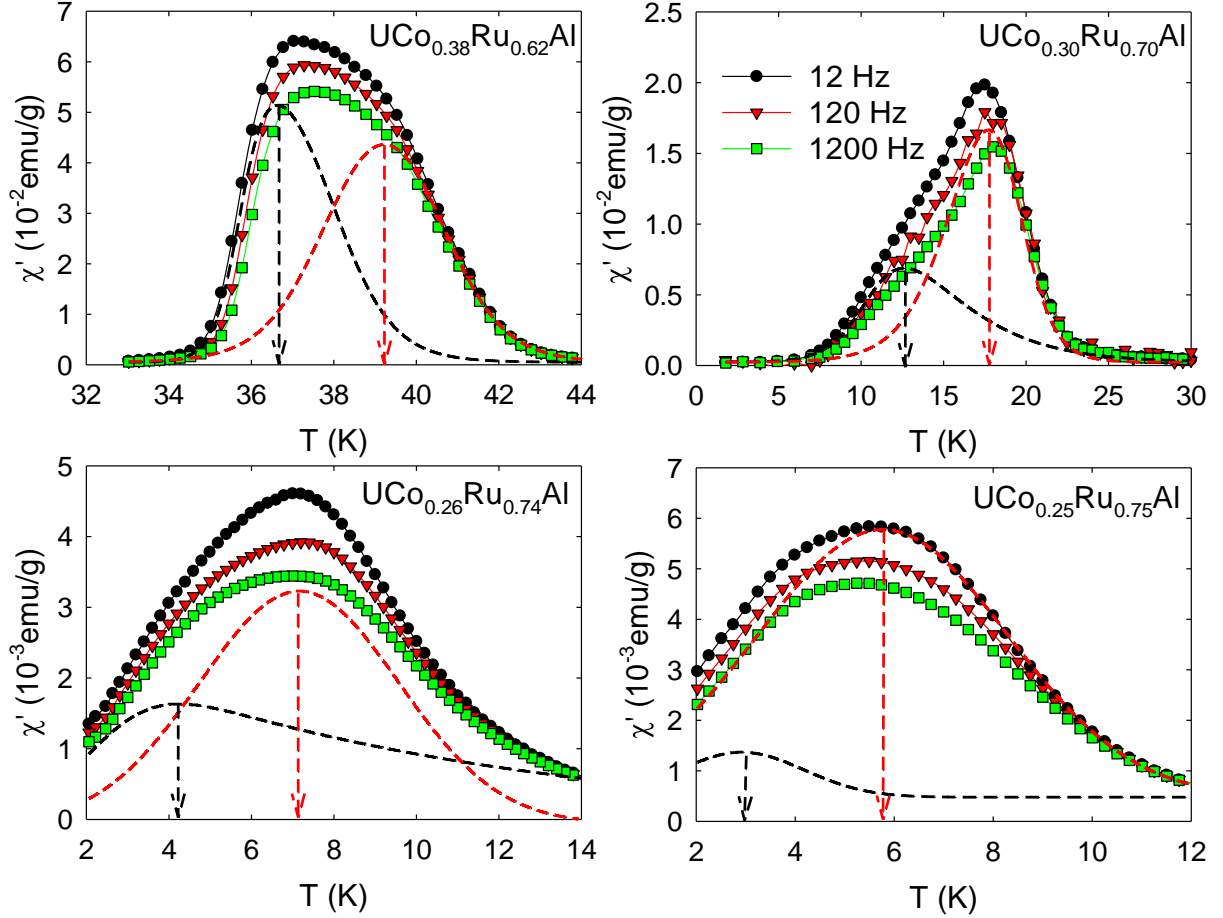

FIG. 3. (Color online) AC susceptibility of UCo$_{1-x}$Ru$_x$Al at various frequencies of the modulation field of 0.3mT applied along the axis *c*. The dashed lines represent fits of the data measured at the frequency 12 Hz.

The temperature dependence of the magnetization was measured in various magnetic fields with no sign of influence of a FM impurity. Selected results of the inverse susceptibility are shown in Fig. 4. The data were analyzed using a modified Curie Weiss law (1), which accounts well for the inverse susceptibilities being strongly nonlinear in the PM region.

$$\chi = \chi_0 + \frac{C}{T - \theta_P} \quad (1)$$

Good agreement holds down to 50 K, where deviations appear due to the proximity of the magnetically ordered states.

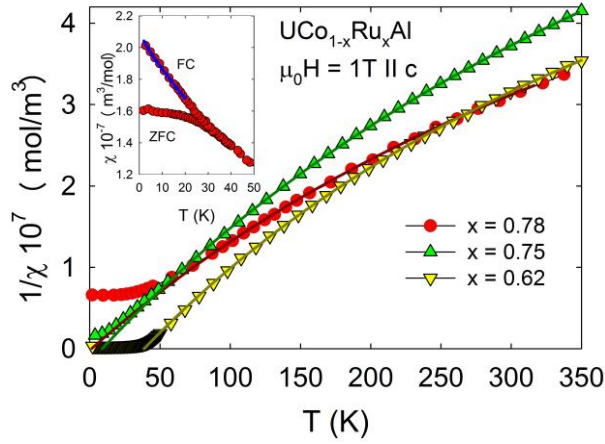

FIG. 4. (Color online) Inverse susceptibilities of the selected compounds. Inset shows $\chi$ of $UCo_{0.22}Ru_{0.78}Al$, blue line represents $\chi \approx \chi_0 + T^{-\gamma}$ fit.

The systematic decrease of $\theta_P$ with increasing $x$ closely follows the gradual decay of the FM. This result is in good agreement with the original work[12]. Finally, we have fit $\chi$ to the power law $\chi \approx \chi_0 + T^{-\gamma}$. The result is displayed in Fig.4 and will be discussed latter.

The magnetization loops (Fig. 5) confirmed the strong magnetocrystalline anisotropy of all the compounds, including PM $UCo_{0.22}Ru_{0.78}Al$. The magnetization in the basal plane is very close to Pauli PM behavior and essentially independent of the c-axis features.

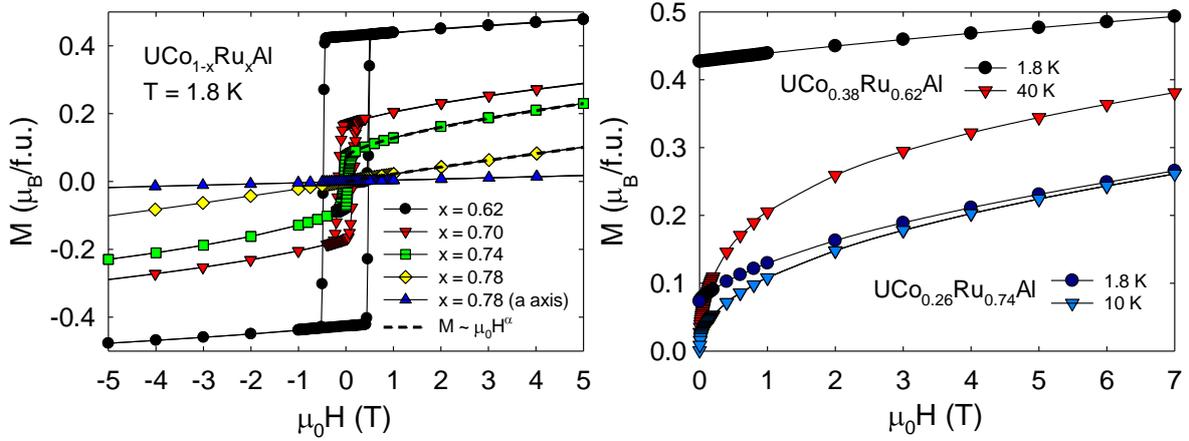

FIG. 5. (Color online) Left) Magnetization loops of all prepared $UCo_{1-x}Ru_xAl$ single crystals in magnetic fields applied along the c-axis with an exception mentioned in the figure. Dashed lines represent $M \approx H^\alpha$ fits. Right) Magnetization loops of $UCo_{0.38}Ru_{0.62}Al$ and $UCo_{0.26}Ru_{0.74}Al$ at 1.8 K and just above $T_C$ are shown.

$\mu_{spont.}$ and $\mu_{sat.}$ decrease with increasing $x$ (Table III). We have tested scaling of the magnetization isotherms in the vicinity of $x_{crit.}$ to $M \approx \mu_0 H^\alpha$, the form predicted for disordered systems[34]. Power law fits are displayed in Fig. 5 and will be discussed later.

Another distinct feature of the $UCo_{1-x}Ru_xAl$ alloys is the evolution of the magnetization loops as a function of temperature (Fig. 5). $\mu_{spont.}$ of $UCo_{0.38}Ru_{0.62}Al$ remains almost unchanged up to 30 K but drops rapidly approaching $T_C$. Similarly magnetization loops also lose their originally rectangular shape to more magnetic soft behavior. The $\mu_{spont.}$ of the other alloys is much reduced even at low temperatures which leads to reduced magnetic entropy $S_{mag}$. The values of $S_{mag.}$, determined from the magnetization isotherms using (2), are listed later in conclusions.

$$\Delta S_{mag} = \int_0^H \left(\frac{\partial M(H,T)}{\partial T}\right)_H dH \quad (2)$$

### 3.3 Polarized neutron diffraction

We have performed a polarized neutron diffraction (PND) study of $UCo_{0.38}Ru_{0.62}Al$ to understand the FM phase in the $UCo_{1-x}Ru_xAl$ system at the microscopic level. We first accurately determined the crystal structure (Table II) since this technique gives access to the ratio between the magnetic and nuclear structure factors. PND was carried out at 4 K and in a 4 T magnetic field applied along the crystallographic $c$ axis. Spin densities were deduced from the measured magnetic structure factors through a maximum entropy reconstruction[35-37]. The unit cell was divided into 50 x 50 x 50 = 125000 smaller cells before computation of the distribution of magnetic moment. The reconstruction was started from a flat magnetization distribution with a total moment in the unit cell equal to the bulk magnetization measured at the same temperature and magnetic field. The resulting distribution (magnetic density with the highest probability), projected onto the basal plane, is shown in Fig. 6. The U ions evidently carry the major part of the magnetic moment. Clouds of rather delocalized magnetization density exist in the interstitial spaces. We could not use common spherical integration to estimate the sizes of the magnetic moments from the reconstructed density considering the extended magnetization rather far away from the U centers and especially $T$-sites. We used the (spherical) dipole approximation to estimate the spin and orbital components of the magnetic moments carried by the U and Co ions, similar to the procedure used for $UCo_{1-x}Ru_xGe$[38] using the FullProf[29]/WinPlotr[28] software.

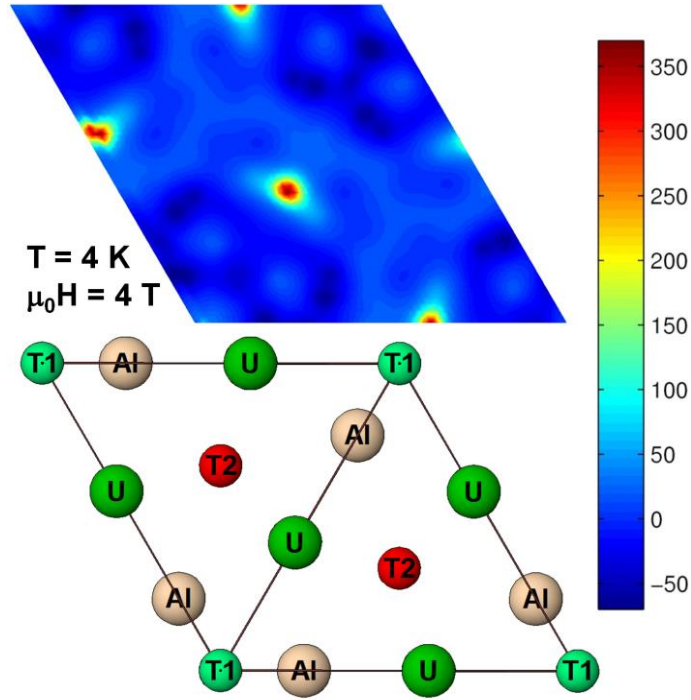

FIG. 6. (Color online) Magnetization density map of the $UCo_{0.38}Ru_{0.62}Al$ in basal plane. The red shade represents the atoms in the T-Al layer, green shade U-T layer. The scale of the map is in $m\mu_B$ units.

The spherical integrals for both possible states of uranium ($U^{3+}$ and $U^{4+}$)[39] are similar, which disqualifies this method for the determination of the valence of the U ion. For our refinement we wrote the $U^{3+}$ form factor in the form $f_m(s) = W_0 \langle j_0(s) \rangle + W_2 \langle j_2(s) \rangle$, where $s = \sin\theta / \lambda$ is the scattering vector. Tabulated approximations of the $\langle j_0(s) \rangle$ and $\langle j_2(s) \rangle$ functions are taken from Ref.[40]. $W_0$ and $W_2$ are the fitted parameters related to the spin and orbital moments: $\mu_L = W_2$ and $\mu_S = W_0 - W_2$. For the weaker moments on the Co ions we considered only the spin part. The results are summarized in Table IV.

TABLE IV. Components of the magnetic moment on the U and Co(Ru) positions from the refinement of the polarized neutron diffraction data. All values are displayed in $\mu_B$ unit. The value of $\mu_{tot.}$ was calculated as $\mu_L^U + \mu_S^U + \mu_S^{Co1} + \mu_S^{Co2}$. The value of $\mu_{int.}$ was calculated as $\mu_{bulk} - \mu_{tot.}$. The table also gives the free U ion and parent compounds values for comparison. $\mu_{bulk}$ represents the value of the macroscopic magnetization at the conditions of the PND experiment.

| composition | $\mu_{bulk}$ | $\mu_L^U$ | $\mu_S^U$ | $\mu_{tot.}^U$ | $\mu_S^{Co1}$(U) | $\mu_S^{Co2}$(Al) | $\mu_{tot.}$ | $|\mu_L^U/\mu_S^U|$ | $\mu_{int.}$ |
|---|---|---|---|---|---|---|---|---|---|
| $x = 0.62$ | 0.42 | 0.611(48) | -0.376(11) | 0.235(37) | 0.110(25) | 0.045(7) | 0.390(68) | 1.63(12) | 0.030 |
| $U^{3+}$ free ion[41] | | 5.585 | -2.169 | 3.416 | | | | 2.6 | |
| $U^{4+}$ free ion[41] | | 4.716 | -1.432 | 3.284 | | | | 3.3 | |

On the uranium site, the orbital moment $\mu_L^U$ is the leading part and parallel to the applied magnetic field. The weaker spin moment $\mu_S^U$, is coupled antiparallel to $\mu_L^U$, similar to what has been observed in many uranium-based compounds[42]. The observed $\mu_L^U$ and $\mu_S^U$ are much weaker than the free ions values. Considering the very weak moments on the $T$ sites, any attempt at introducing extra parameters (e.g. a Ru form factor) failed. Neutron diffraction data in relation to macroscopic data will be discussed later in detail.

### 3.4 Heat capacity

We have measured the temperature dependence of the heat capacity for all the compounds (Fig. 7). Surprisingly, a tiny anomaly at $T_C$ was observed only in UCo$_{0.38}$Ru$_{0.62}$Al even though the magnetization study revealed FM order for all $x < 0.77$.

We used the same procedure as in Ref.[43] to evaluate the magnetic part $C_{mag}$ and related magnetic entropy $S_{mag}$ in UCo$_{0.38}$Ru$_{0.62}$Al. We obtained a Sommerfeld coefficient $\gamma = 55$ mJ/molK$^2$. The maximum of $C_{mag}$ (2.5 J/molK) coincides well with the position of the anomaly at $T_C = 35$ K. The related magnetic entropy $S_{mag}$ saturates at value 0.17Rln2. For the phonon part we used a model with 3 acoustic branches, with a Debye temperature $\theta_D = 155$ K and two 3-fold degenerate optical branches with $\theta_{E1} = 215$ K and $\theta_{E1} = 650$ K. The model is in reasonable agreement with Ref.[43]. The already weak anomaly is suppressed by a magnetic field of 1 T.

For UCo$_{0.30}$Ru$_{0.70}$Al, UCo$_{0.26}$Ru$_{0.74}$Al, UCo$_{0.25}$Ru$_{0.75}$Al and UCo$_{0.22}$Ru$_{0.78}$Al we evaluated the phonon part with parameters: $\theta_D = 150$ K 3-fold degenerated $\theta_{E1} = 185$ K and $\theta_{E1} = 650$ K and obtained the $\gamma$ coefficients of 65-70 mJ/molK$^2$ – see Table III. Considering the $\gamma$ value of UCo$_{0.38}$Ru$_{0.62}$Al and $\gamma_{URuAl} = 45$ mJ/molK$^2$ [44] one sees a weak broad maxima in $\gamma$ evolution around $x_{crit.}$.

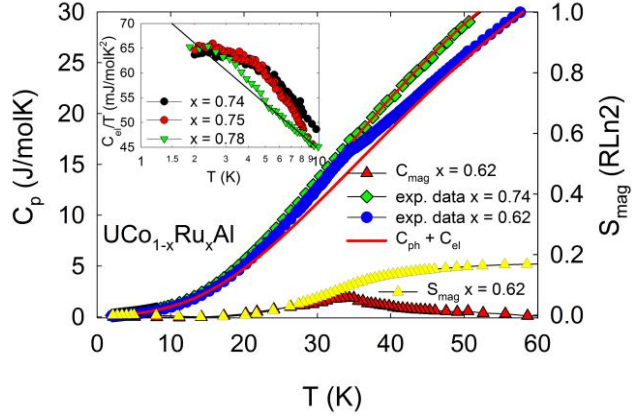

FIG. 7. (Color online) Temperature dependence of the UCo$_{0.38}$Ru$_{0.62}$Al and UCo$_{0.26}$Ru$_{0.74}$Al heat capacities. The UCo$_{0.38}$Ru$_{0.62}$Al data are decomposed into a phonon $C_{ph}$ and a magnetic $C_{mag}$ contribution. The right axis represents the related magnetic entropy $S_{mag}$ The inset shows $C_{el}/T$ in log$T$ scale; solid line represents $C/T \propto T^{-\gamma}$ ($\gamma = 0.03$).

Since a non-Fermi liquid state develops in the neighboring UCo$_{1-x}$Ru$_x$Ge[24] and URh$_{1-x}$Ru$_x$Ge[25] systems, we have analyzed the heat capacity within this scenario[45, 46] and plotted the data as $C_{el}(T)/T$ on a log$T$ scale ($C_{el} = C_{exp} - C_{ph}$). The curvature changes with increasing Ru content. UCo$_{0.22}$Ru$_{0.78}$Al in the vicinity of $x_{crit.}$ approaches a linear trend but fully linear behavior as in URh$_{1-x}$Ru$_x$Ge[25] was not reached. In comparison to UCo$_{1-x}$Ru$_x$Ge[24] we have detected only weak enhancement of the Sommerfeld coefficient in the vicinity of $x_{crit.}$. We have also evaluated the low temperature data of UCo$_{0.22}$Ru$_{0.78}$Al as $C/T \propto T^{-\gamma}$ according to the model for disordered systems with $\gamma = 0.03$ in temperature interval 5 – 10 K (Fig. 7).

### 3.5 Electrical and Hall resistivity

The temperature dependence of the electrical resistivity of UCo$_{0.38}$Ru$_{0.62}$Al (Fig. 8) shows a cusp at ~ 39 K corresponding to $T_C$. Below 39 K the electrical resistivity follows $\rho = \rho_0 + AT^2$ behavior with $\rho_0 = 23.6$ μΩcm and $A = 0.004$ μΩcmK$^{-2}$. A field of 1 T applied along the c axis is enough to smear out the anomaly consistent with the heat capacity observations. Similarly, there is no sign of any anomaly in the compounds with $x > 0.6$ and the trend at low temperature does not follow ~$T^2$ behavior anymore. Instead, $\rho = \rho_0 + AT^n$ with $\rho_0 = 17.63$ μΩcm, $A = 0.010$ μΩcm/K$^{1.7}$ and $n = 1.7$ for UCo$_{0.30}$Ru$_{0.70}$Al, $\rho_0 = 37.12$ μΩcm, $A = 0.020$ μΩcm/K$^{1.48}$ and $n = 1.48$ for UCo$_{0.26}$Ru$_{0.74}$Al, $\rho_0 = 23.3$ μΩcm, $A = 0.018$ μΩcm/K$^{1.4}$ and $n = 1.4$ for UCo$_{0.25}$Ru$_{0.75}$Al and $\rho_0 = 1.2$ mΩ, $A = 0.001$ μΩ/K$^{5/3}$ and $n = 5/3$ for UCo$_{0.22}$Ru$_{0.78}$Al give reasonable agreement which points to the development of spin fluctuations.

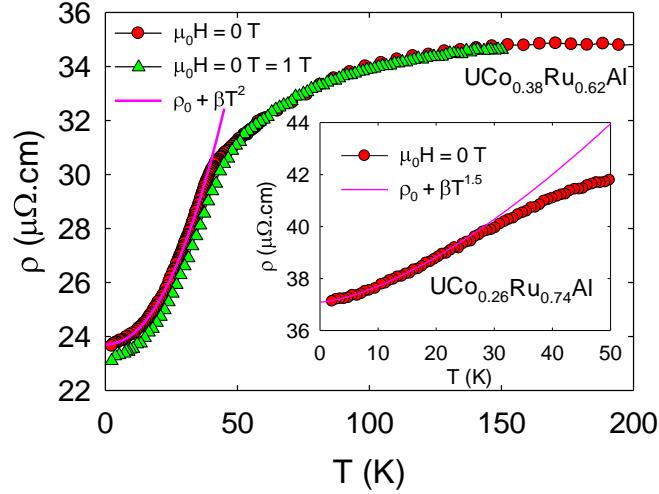

FIG. 8. (Color online) Temperature dependence of the electrical resistivity of UCo$_{0.38}$Ru$_{0.62}$Al in various magnetic fields. The measurements were done in a μ$_0$H ∥ I ∥ c arrangement. The inset shows the behavior of UCo$_{0.26}$Ru$_{0.74}$Al close to $x_{crit.}$.

The Hall effect was measured with the magnetic field applied along the axis $c$ (Fig. 9). The magnetic field polarity was changed by rotating the sample by 180° on a rotator in zero magnetic field. Hysteresis appeared consistently with magnetization loops. At the lowest temperatures, it decreases from 1 T in UCo$_{0.38}$Ru$_{0.62}$Al to 0.22 T in UCo$_{0.26}$Ru$_{0.74}$Al and 0.12 T for UCo$_{0.25}$Ru$_{0.75}$Al, signaling the rapid vanishing of the FM phase. Only UCo$_{0.38}$Ru$_{0.62}$Al showed the ideal rectangular shape at 0.03K.

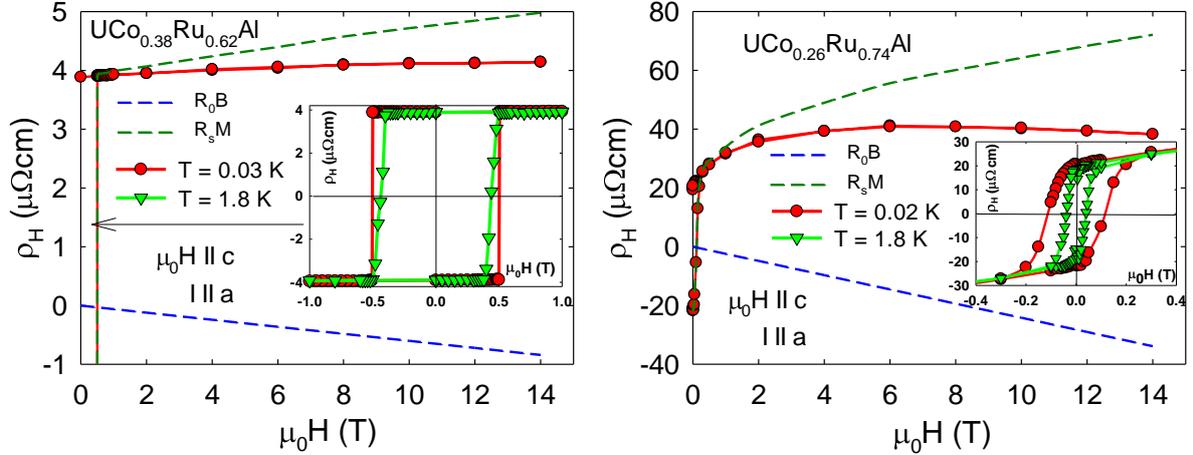

FIG. 9. (Color online) (Left) Magnetic field dependence of the Hall Effect measured in UCo$_{0.38}$Ru$_{0.62}$Al single crystal at 1.8 and 0.03 K. The values at 1.8 and 0.03 K roughly overlap except for the low temperature part: the hysteresis is ideally rectangular at low temperature. See the inset of the figure. (Right) Magnetic field dependence of the Hall Effect measured in UCo$_{0.26}$Ru$_{0.74}$Al at 1.8 and 0.02 K and magnetic field up to 14 T.

In magnetic materials, $\rho_H$ may be empirically described as

$\rho_H(B) = R_0 B + R_S M$ (Eq.3)

where $R_0$ and $R_S$ are the ordinary and the anomalous Hall coefficients. At high field in UCo$_{0.38}$Ru$_{0.62}$Al, we obtained $R_0$ = -0.06 μΩcm/T and $R_S$ = 9.1 μΩcm/T. In UCo$_{0.26}$Ru$_{0.74}$Al, $R_0$ = -2.4 μΩcm/T and $R_S$ = 250 μΩcm/T and in UCo$_{0.25}$Ru$_{0.75}$Al, $R_0$ = -1.9 μΩcm/T, $R_S$ = 330 µΩcm/T. In all cases, the majority of the Hall resistance comes from the anomalous Hall effect (AHE), while the ordinary Hall effect (OHE) is negative, opposite to the parent UCoAl[2, 10].

The enhancement of the AHE for $x = 0.74$ and $0.75$ is the most striking effect. Although URuAl[7] was studied in detail, there is no related Hall effect measurement. Generally $\rho_H$ is a very complex quantity in heavy fermion multiband systems. We suggest that the change of the effective mass of carriers has a strong influence on $\rho_H$. It can be deduced from inelastic $A$ coefficient of the $AT^n$ electrical resistivity term[2]. In agreement the $A$ coefficient continuously grows as a function $x$ up to $A = 0.063$ µΩcm/K$^2$ for parent URuAl[7] but the strong enhancement typical for systems with QCP was not observed[47]. On the other hand, the value of $A$ in URuAl is still considerably lower than that of well-established heavy fermion compounds such as UPt$_3$[48]. Thus, we consider a second scenario taking into account possible variation of the effective carrier number upon substitution due to one electron less of Ru.

**4 Discussion**

We have successfully performed detailed investigations of the Ru rich substituted UCoAl compounds by a series of macroscopic and microscopic methods.

The main parameters of the anomalously developed FM state are summarized in Table V. The spontaneous magnetization develops exclusively along the axis $c$. Comparing with other U$TX$ compounds, we note that for UCo$_{0.38}$Ru$_{0.62}$Al, closest to the maximum in the FM dome, the value of $T_C$ is very close to that of UPtAl[43] and URhAl[49]. The spontaneous magnetic moment $\mu_{spont.}$, however, is much reduced, corresponding to only ~30 % of the moment in UPtAl[43] and ~40 % of that in URhAl[49]. It is in fact closer to the moment of parent UCoAl at the metamagnetic transition[8]. The value of the effective moment $\mu_{eff.}$ is almost constant throughout the studied range of concentrations and lower than in UPtAl (2.8 µ$_B$/f.u.)[50] considered as an itinerant system, and significantly lower than expected for the free U$^{3+}$ and U$^{4+}$ ions. The reduced $\mu_{eff.}$ here reflects the high delocalization of the 5$f$ states in the UCo$_{1-x}$Ru$_x$Al in agreement with the extremely low values of $S_{mag}$. Consequent very weak or missing anomalies in the heat capacities and electrical resistivity are evidence of the U itinerant ferromagnetism[51].

Key results are provided by PND, showing the development of the hybridization strength through the system. Microscopic evidence for delocalization of the 5$f$ states can be deduced from the value of the $|\mu_L^U/\mu_S^U|$ ratio. Generally, a higher value of $|\mu_L^U/\mu_S^U|$ suggests a stronger localization of the 5$f$ electrons and vice versa[38, 52, 53]. Here, in the vicinity of $T_{C,max}$ for UCo$_{0.38}$Ru$_{0.62}$Al, $|\mu_L^U/\mu_S^U|$ equals only 1.63(12) which is far from the U$^{3+}$ and U$^{4+}$ free ion values[38] and comparable to that seen in the generally accepted itinerant FM UCoGe[38, 54, 55].

TABLE V Comparison of the lattice parameters, d$_{(U-U)}$ distance and magnetic parameters of UCo$_{1-x}$Ru$_x$Al with equivalent parameters of isostructural UTAl FMs.

| | T$_C$ (K) | $\mu_{spont.}$ ($\mu_B$/f.u.) | T$_C$/$\mu_{spont.}$ | S$_{mag}$ (J/molK) | a (Å) | c (Å) | a/c | d$_{(U-U)}$ (Å) |
|---|---|---|---|---|---|---|---|---|
| UPtAl | 44[43] | 1.2[43] | 36.7 | 4.1[43] | 7.012[56] | 4.127[56] | 0.588 | 3.6[56] |
| UIrAl | 64[57] | 0.96[57] | 66.7 | 1.9[58] | 6.95[57] | 4.01[57] | 0.576 | |
| URhAl | 27[59] | 0.94[59] | 28.7 | 2.3[60] | 6.965[56] | 4.019[56] | 0.577 | 3.63[56] |
| UCoAl | - | - | - | - | 6.686[56] | 3.966[56] | 0.593 | 3.5[56] |
| UCo$_{0.99}$Ru$_{0.01}$Al | 20[15] | 0.38[14] | 58.8 | | | | | |
| UCo$_{0.38}$Ru$_{0.62}$Al | 38.5 | 0.44 | 86.4 | 1.6 | 6.856 | 3.969 | 0.578 | 3.57 |
| UCo$_{0.25}$Ru$_{0.75}$Al | 7.5 | 0.04 | 187.5 | 0.1 | 6.869 | 3.988 | 0.580 | 3.57 |
| URuAl | - | - | - | - | 6.895[56] | 4.029[56] | 0.584 | 3.6[56] |

$|\mu_L^U/\mu_S^U| \approx 2.3$ was found for the field induced FM state of UCoAl[61]. The URuAl magnetic ground state is rather unusual and was the subject of theoretical calculations. Calculated band structure based on spin and orbital-polarized LSDA method supposes that the PM state is caused by almost perfect cancelation of the reduced $\mu_L^U$ and $\mu_S^U$ resulting in a $|\mu_L^U/\mu_S^U|$ ratio of only 1.3[7]. Here is a discrepancy between the theoretical and the experimental $|\mu_L^U/\mu_S^U| \approx 2.5$[62] which is, however, so far the highest value and not having any correspondence to all other UTAl[63]; see the ratios of the neighboring itinerant FM URhAl[59] and UPtAl[64]. One must consider the PND detection limit in relation to the extremely weak magnetic moment of the PM URuAl induced by magnetic field. $|\mu_L^U/\mu_S^U| = 1.63$ of UCo$_{0.38}$Ru$_{0.62}$Al is in between that of UCoAl and that expected for URuAl and thereby confirms the suggested scenario for the anomalous PM of the later compound.

The strength of the hybridization can be deduced from the proportion of the induced magnetic moments in T - sites. While almost equivalent induced magnetic moments were found in Co sites in UCoAl[61] the Ru sites in URuAl are magnetically inequivalent with significantly larger magnetic moment in the U - T plane[62]. Our results find two times higher induced magnetic moment for T site in the U - T plane. This gives evidence for growth of the hybridization strength in U - T plane most likely driving the magnetism in UCo$_{1-x}$Ru$_x$Al. It is in agreement with the observation by Hall effect. $\rho_H$ of UCo$_{0.38}$Ru$_{0.62}$Al is similar to UCoAl at the metamagnetic transition[2]. Significantly enhanced values of $\rho_H$ were detected for UCo$_{0.25}$Ru$_{0.75}$Al and UCo$_{0.24}$Ru$_{0.76}$Al, which agree well with the suggested importance of a number of seven d electrons for the paramagnetic state of URuAl[7]. It is one less than in UCoAl and causes stronger charge carrier transfer from the U dominantly via hybridization in the U - T plane. For x > 0.6 hybridization rapidly changes and approaches to that of URuAl. On the other hand, only weak development of the T$_C$/$\mu_{spont.}$ ratios for x < 0.62 (inset of Fig. 10) supports that the hybridization strength between U and both T sites changes only smoothly in this region. Generally, we suppose that the hybridization in UCo$_{1-x}$Ru$_x$Al plays the role of a mediator of the unusually strong indirect interaction although the 5f states remain considerably delocalized. Finally, we did not find any scaling of the magnetic parameters in the ZrNiAl-type UTAl FMs representing more complex magnetic interactions in comparison to neighboring TiNiSi-type UTX FMs (with one unique T site) where scaling of S$_{mag}$ and $|\mu_L^U/\mu_S^U|$ were found for a large series of pure and also substituted compounds as the functions of d(U-U) and T$_C$[24, 38].

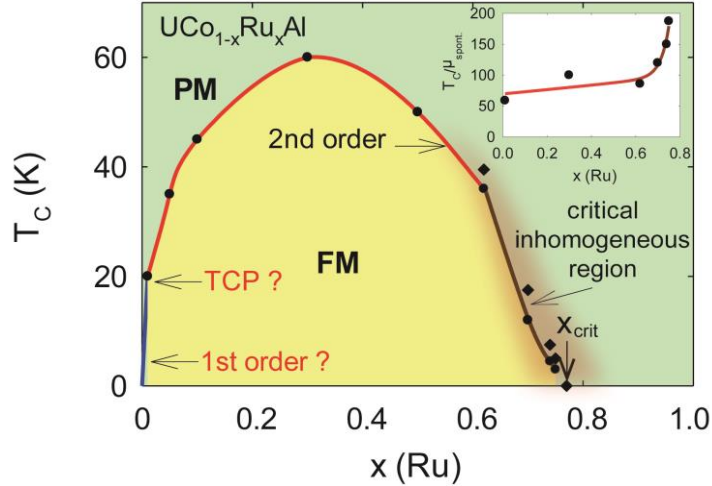

FIG. 10 (Color online) Magnetic phase diagram of the $UCo_{1-x}Ru_xAl$ system. The phase diagram was constructed on the basis of ref.[13] ($0 < x < 0.01$), ref.[12] ($0.01 < x < 0.6$) and our AC susceptibility data ($0.62 > x > 0.78$). The inset shows evolution of the $T_C/\mu_{spont.}$ ratio as a function of $x$; line is guide for the eye. $T_C/\mu_{spont.}$ ratio for $x = 0.01$ was calculated using the data available in ref.[13], ref.[12] was used for $x = 0.3$.

We have estimated $x_{crit.} \approx 0.77$ where FM phase vanishes (Fig. 10). Our previous conclusions show that hybridization strength develops gradually with increasing $x$. In contrast, FM dome is markedly asymmetric at the boundaries. More rapid change of $T_C$ (~20 K / 1 % Ru) is seen on the UCoAl side[13, 14] than that (~ -3 K / 1 % Ru) on the URuAl side. It is known that in clean FMs original second order transition transforms to first order at TCP by tuning of the critical parameter[27]. It was recently confirmed in FM URhAl[65]. With increasing disorder, the temperature of the TCP decreases and above a critical disorder strength a QCP is realized in zero temperature. In agreement, Ru (and many other $T$ metals) substitution causes instantaneous transformation of the metamagnetic UCoAl to the FM state[14] and transition changes from first to second order. Then, FM is established at the temperature where the original spin fluctuations appear. This scenario was confirmed by NMR in neighboring $UCo_{1-x}Fe_xAl$[5, 22, 66]. We expect existence of a TCP in $UCo_{1-x}Ru_xAl$ around 1% of Ru (Fig.10).

On the opposite side of the FM dome near $x_{crit.}$, there is no evidence for the first order transition and the electrical resistivity exponent drops from $\sim T^{2.0}$ to $\sim T^{1.5}$ together with straightening of the $C_{mag}/T$ on logarithmic scale. It is evidence for approaching a QCP[23-25]. High $x$ $UCo_{1-x}Ru_xAl$ alloys, however, show apparent differences from the critical behavior of similar alloy systems[23-25]. Enhancement of the Sommerfeld $\gamma$ coefficient is broad and very weak around $x_{crit.}$ and electrical resistivity starts to vary $\sim T^2$ already when $T_C$ is still very high (18 K of $UCo_{0.30}Ru_{0.70}Al$). The wide region of electrical resistivity exponent variation suggests that a finite critical region develops instead of a thermodynamic singularity.

Notable broadening of the critical region is predicted for Griffiths effect where a magnetically inhomogeneous phase develops in the disordered systems and FM/PM boundary smears due to so-called rare region[67, 68] (Fig. 10). These rare spatial regions are locally magnetic while the bulk is nonmagnetic. Characteristic parameters of the locally magnetic regions are their volume $V$ depending on disorder strength and characteristic energy $\varepsilon$ exponentially depending on $V$. Final energy spectrum is defined as $P(\varepsilon) \sim \varepsilon^{\lambda-1}$ where $\lambda$ is the Griffiths exponent of which power law spectrum gives singularity in many quantities[33]. Signature of the quantum Griffiths phase was studied and confirmed only in a few FM systems such as $CePd_{1-x}Rh_x$[69, 70] and particularly $Ni_{1-x}V_x$ and $Ni_{1-x}Cu_x$ where weak itinerant FM is suppressed by substitution[33, 34, 71, 72]. Data analysis finds that $UCo_{1-x}Ru_xAl$ features also fit the Griffiths scenario. Magnetic clusters within the rare region respond to the frequency of the modulation field in AC susceptibility together with deviation of the magnetization measured in FC and ZFC regime which is in agreement with our observations. Magnetization isotherms in Arrott plots show deviation from linear behavior for $x$ close to $x_{crit.}$ and critical exponents develop rapidly. The Griffiths phase scenario predicts scaling of the critical exponents where $M \propto \mu_0 H^\alpha$ with $0 \leq \alpha \leq 1$ and $\chi \propto T^{-\gamma}$ with relation of Griffiths exponent $\lambda = \alpha = 1 - \gamma$ [33]) for $x > x_{crit.}$.

Magnetization isotherms analysis gave $\alpha \approx 0.6$ for $x = 0.74$ and $0.75$ and $\alpha = 0.91$ for $x = 0.78 > x_{crit.}$, respectively. Major difference from $Ni_{1-x}V_x$, is one order higher $T_{C,max.}$ of Ni (630 K) than $T_{C,max.}$ in $UCo_{1-x}Ru_xAl$. Then, the reasonable paramagnetic region for $\gamma$ investigation is reduced. We did not acquire valid results for $x < x_{crit.}$ due to the presence of the FM order in the low temperature. Nevertheless, $\gamma = 0.01$ for $x = 0.78 > x_{crit.}$ was found, which fits well to the Griffiths phase scenario. Existence of magnetic clusters is supported by the $\chi$ data of $UCo_{0.22}Ru_{0.78}Al$. Deviation between FC and ZFC magnetization curves appears although only the PM phase should be present. This resembles observations in the substituted uranium FM US where collapse of the U magnetic moment does not coincide with $T_C$ evolution. μSR clearly demonstrated that the residual magnetic moments in the short range clusters survive in a diluted phase[73]. Substitution in US also causes violation of Vegard's Law for lattice parameters[74] and magnetization loops lose rectangular shape[75].

Local inhomogeneous distribution of Co/Ru can cause a portion of the sample to behave as a diluted PM while the major portion still keeps the character of the bulk FM. In agreement we detected hysteresis by Hall Effect in $UCo_{0.26}Ru_{0.74}Al$ and $UCo_{0.25}Ru_{0.75}Al$ at low temperatures and simultaneously the two peaks in AC susceptibility. The existence of an inhomogeneous region is reflected in the sudden growth of the $T_C/\mu_{spont.}$ ratio close to $x_{crit.}$ (Fig.10). One must consider that the bulk FM phase survives up to very high $x$. 3/4 of the $T$-sites in the vicinity of $x_{crit.}$ are already occupied by Ru creating the PM phase. We suppose from the AC peaks proportion that diluted PM phase can appear from quite low $x \approx 0.6$ where hybridization already has locally the features observed for URuAl, and such a region gradually dominates the sample volume approaching $x_{crit.}$. The inhomogeneous region in the phase diagram in (Fig. 10) is constructed on the basis of AC susceptibility data $\chi'$ which give the strongest evidence for the existence of the PM diluted phase. The detailed boundary will be the subject of further research.

Originally, a non-simple FM/PM transition was deduced for polycrystalline samples based on the development of $\mu_{spont.}$ and hysteresis together with two peaks in $\chi'$ data. The present study shows that even in single crystals $\mu_{spont.}$ emerges several Kelvins prior to hysteresis (Table III) but completely separated double peaks in $\chi'$ were not detected anymore[12]. The simplest explanation of this result would be that better Ru homogeneity of our single crystals at the atomic scale which narrows the inhomogeneous region substantially.

## 5 Conclusions

We have successfully prepared a series of UCoAl single crystals substituted by Ru, constructed the $T-x$ phase diagram and estimated the critical concentration $UCo_{0.23}Ru_{0.77}Al$ where $T_C$ approaches zero. Our research by microscopic and macroscopic methods pointed to the anomalous role of hybridization in the $UCo_{1-x}Ru_xAl$ system, which mediates the strong indirect interaction responsible for high $T_C$. Simultaneously spontaneous magnetic moments remain very low at the magnitude of the metamagnetic transition of UCoAl, also evident from the extremely low $S_{mag}$ for all compounds, causing disappearance of the magnetic anomalies both in electrical resistivity and heat capacity data and very low $|\mu_L^U/\mu_S^U|$ ratio of $UCo_{0.38}Ru_{0.62}Al$. PND detected growth of the hybridization strength between U and $T$ site in the $U-T$ plane with increasing $x$. Simultaneously $|\mu_L^U/\mu_S^U| = 1.63(12)$ of $UCo_{0.38}Ru_{0.62}Al$ is in between that of UCoAl and URuAl and confirms a suggested theoretical scenario for the anomalous PM state of the later compound.

We also conclude that UCo$_{1-x}$Ru$_x$Al represents a unique system with two simultaneous critical regions of different type. Disorder seems to be the key parameter dominating the criticality on both sides. While on the UCoAl side the critical behavior has the character of a clean limit FM metal with first order transition, the criticality on the URuAl side is more complex with the signature of a magnetically inhomogeneous system. Data analysis implies that a finite Griffiths phase scenario is the most likely explanation of the region. The ZrNiAl-type structure and resulting complex magnetic interactions involving 5$f$ electron states however make investigation of the Griffiths phase difficult. We have found that the system behaves inhomogeneously in the narrow temperature interval ~ 3 K already for $x > 0.6$, which makes the FM/PM transitions so indistinct. μSR spectra measurements are highly desirable to verify existence of the magnetic clusters and to define properly the potential Griffiths phase boundary. The existence of the TCP in the UCo$_{1-x}$Ru$_x$Al system at the UCoAl boundary should be confirmed experimentally by NMR method.

## 6 Acknowledgemsents

Authors thank to Z. Fisk, S. Kambe and Mohsen M. Abd-Elmeguid for fruitful discussion of the experimental data. This work was supported by the Czech Science Foundation no. P204/12/P418. Experiments performed in MLTL (see: http://mltl.eu/) were supported within the program of Czech Research Infrastructures (project LM2011025).